\documentclass[%
 reprint,
superscriptaddress,
 amsmath,amssymb,
 aps,
prl,
]{revtex4-2}

\usepackage[T1]{fontenc}
\usepackage[utf8]{inputenc}
\usepackage{mathtools}
\usepackage{graphicx}
\usepackage{xcolor}
\usepackage{enumerate}
\usepackage{wrapfig}
\usepackage{lipsum}  
\usepackage{dcolumn}
\usepackage{braket}
\usepackage{appendix}

\usepackage{amsmath,amsthm,amssymb,amsfonts}
\usepackage{bm}
\usepackage{physics}

\usepackage[colorlinks=True,citecolor=blue,linkcolor=blue,urlcolor=blue]{hyperref}

\begin{document}
\preprint{APS/123-QED}

\title{Stark Tuning and Charge State Control in Individual Telecom C-Band Quantum Dots }

\author{N.J. Martin}
\email{n.j.martin@sheffield.ac.uk}
\author{A.J. Brash}
\author{A. Tomlinson}
\affiliation{School of Mathematical and Physical Sciences, University of Sheffield, Sheffield S3 7RH, UK}%
\author{E.M. Sala}
\affiliation{EPSRC National Epitaxy Facility, University of Sheffield, Sheffield S1 4DE, UK}%
\affiliation{School of Electrical and Electronic Engineering, The University of Sheffield, North Campus, Broad Lane, S3 7HQ Sheffield, UK}%

\author{E.O. Mills}
\affiliation{School of Mathematical and Physical Sciences, University of Sheffield, Sheffield S3 7RH, UK}%
\author{C.L. Phillips}
\affiliation{School of Mathematical and Physical Sciences, University of Sheffield, Sheffield S3 7RH, UK}%
\author{R. Dost}
\affiliation{School of Mathematical and Physical Sciences, University of Sheffield, Sheffield S3 7RH, UK}%
\author{L. Hallacy}
\affiliation{School of Mathematical and Physical Sciences, University of Sheffield, Sheffield S3 7RH, UK}%
\author{P. Millington-Hotze}
\affiliation{School of Mathematical and Physical Sciences, University of Sheffield, Sheffield S3 7RH, UK}%
\author{D. Hallett}
\affiliation{School of Mathematical and Physical Sciences, University of Sheffield, Sheffield S3 7RH, UK}%
\author{K.A. O'Flaherty}
\affiliation{EPSRC National Epitaxy Facility, University of Sheffield, Sheffield S1 4DE, UK}%
\affiliation{School of Electrical and Electronic Engineering, The University of Sheffield, North Campus, Broad Lane, S3 7HQ Sheffield, UK}%
\author{J. Heffernan}
\affiliation{EPSRC National Epitaxy Facility, University of Sheffield, Sheffield S1 4DE, UK}%
\affiliation{School of Electrical and Electronic Engineering, The University of Sheffield, North Campus, Broad Lane, S3 7HQ Sheffield, UK}%
\author{M.S.~Skolnick}
\author{A.M Fox}
\author{L.R. Wilson}
\affiliation{School of Mathematical and Physical Sciences, University of Sheffield, Sheffield S3 7RH, UK}%

\begin{abstract} 
Telecom-wavelength quantum dots (QDs) are emerging as a promising solution for generating deterministic single-photons compatible with existing fiber-optic infrastructure. Emission in the low-loss C-band minimizes transmission losses, making them ideal for long-distance quantum communication. In this work, we present a demonstration of both Stark tuning and charge state control of individual InAs/InP QDs operating within the telecom C-band. These QDs are grown by droplet epitaxy and embedded in a InP-based $n^{++}$--$i$--$n^{+}$ heterostructure, fabricated using MOVPE. The gated architecture enables the tuning of emission energy via the quantum confined Stark effect, with a tuning range exceeding 2.4~nm. It also allows for control over the QD charge occupancy, enabling access to multiple discrete excitonic states. Electrical tuning of the fine-structure splitting is further demonstrated, opening a route to entangled-photon-pair generation at telecom wavelengths. The single-photon character is confirmed via second-order correlation measurements. These advances enable QDs to be tuned into resonance with other systems, such as cavity modes and emitters, marking a critical step toward scalable, fiber-compatible quantum photonic devices.
\end{abstract}

\maketitle

Photonic quantum technologies promise advances in communication, computation, and sensing, with applications including QKD, linear-optical quantum computing, and entangled-photon microscopy~\cite{heindel2023, vajner2022, arakawa2020, singh2024, preskill2023, Defienne2024}. For the realisation of these technologies, high-quality single-photon sources are essential, with semiconductor quantum dots (QDs) a leading platform, providing deterministic emission with high purity and strong coherence and near-unity indistinguishability  ~\cite{arakawa2020, michler2009, cao2019, schweickert2018, michler2024, gyger2022}. Moreover, QDs can be engineered for the telecom bands, enabling long-distance fiber integration with minimal loss~\cite{paul2017, olbrich2017, takemoto2015, miyazawa2016, wakileh2024,muller2018,Jaffal2019,haffouz2018,Bucci2024,wakileh2025}; in particular, the C-band (1530–1565 nm) coincides with the fiber-loss minimum of $\sim$0.18–0.20 dB km$^{-1}$ (down to 0.15 dB km$^{-1}$ in ultra-low-loss spools~\cite{cao2019}), making it attractive for practical deployments. Another key benefit to QDs, demonstrated extensively at shorter wavelengths, is electrical Stark tuning~\cite{Pedersen2020,hallett2018electrical,Nowak2014}, which provides precise, in situ spectral control beyond wafer level band selection. This  control enables  spectral alignment to cavities~\cite{Laucht_2009,purcell_brash}, other emitters~\cite{controlling_coherence}, and narrow atomic transitions~\cite{ORCA}. However, despite these motivations, robust Stark tuning and charge-state control of single QDs at the telecom C-band, has proved difficult to realize. In this Letter, we show that these difficulties can be overcome by using a unipolar  \(n^{++}\!-\!i\!-\!n^{+}\) structure.

In the telecom C-band, QDs have already demonstrated competitive results as single-photon sources \cite{joos2024, ge2024, holewa2024, kim2025,nawrath2019,ha2020} and entangled-photon sources \cite{laccotripes2023, anderson2020} in un-gated wafers. However,  the performance of un-gated QDs is often limited by charge noise and spectral diffusion, where in the absence of an applied electric field, fluctuating charges within the semiconductor matrix induce local electric field variations, leading to spectral wandering and broadening that degrade coherence and so indistinguishability~\cite{anderson2021, phillips2024, wells2023, kaupp2023}.To address these challenges, previous work has explored strategies to reduce charge instability by modifying both the growth mode of the QDs and embedding them in doped heterostructures. Charge noise effects may be influenced by the presence of a wetting layer (such as those present in QDs grown using the Stranski-Krastanov (SK) method) which can act as a reservoir for excess carriers and contribute to spectral fluctuations~\cite{skiba2017, anderson2021}. In contrast, with droplet epitaxy (DE), the ‘quasi-wetting’ layer can be engineered and its continuity, thickness, and composition tuned via crystallization temperature \cite{sala_22}. This has led to DE being proposed as a route toward mitigating charge noise and improving spectral stability and therefore is the approach adopted in this work  ~\cite{skiba2017, sala2020, sala2021, sala2022, sala2024,holewa2022DE}.

In addition to growth optimization, gated QD structures embedded in p–i–n diodes or Schottky junctions \cite{NRC2008PRB,NRC2009APL} have also been developed to stabilize the charge environment, suppress spectral diffusion, and enhance photon indistinguishability~\cite{doi:10.1126/sciadv.abc8268,somaschi2016,muller2018,holewa2024,vajner2024,anderson2020}. However, in InP-based structures, p-dopant diffusion is a significant issue; unlike GaAs, where low-diffusivity carbon acceptors enable stable p–i–n devices \cite{Pedersen2020}. InP in MOVPE growth relies on Zn, which has high diffusivity . This can lead to an unintended shrinking of the intrinsic region and non-uniform electric fields. This can significantly limit the effectiveness of p-i-n gating strategies in InP-based quantum photonic devices, and may prevent the creation of the thin membranes \cite{toshiba_tune} required for low mode volume cavities and single mode operation within slab waveguides. Additionally, a drawback of p-type dopants is that the low mobility of holes, which is approximately twenty times lower than that of electrons, can potentially limit high-frequency electrical modulation in p-i-n structures \cite{900_nin, Pedersen2020}. Moreover, p-type doping introduces roughly twice as much free carrier absorption compared to n-type doping\cite{900_nin, absorption}, further degrading device performance, such as decreasing cavity quality factors and increasing transmission losses in waveguides. Recently, similar challenges associated with p-type dopants in GaAs have motivated the use of n-type only doped structures grown by molecular beam epitaxy (MBE) to tune and control the charge state of near infrared emitting InAs/GaAs QDs \cite{900_nin}.

In this work, we introduce an InP \(n^{++}\!-\!i\!-\!n^{+}\) heterostructure incorporating MOVPE-grown droplet-epitaxy InAs/InP QDs and Al\(_{0.48}\)In\(_{0.52}\)As barrier layers. The barriers confine carriers and suppress leakage, allowing a well-defined vertical field across the intrinsic region while maintaining low dark current in the tuning range (see supplementary section S3). This enables meV-scale Stark tuning ($> 2.4~\mathrm{nm}$), deterministic charge-state control, control over the fine structure splitting (FSS)   and a substantial reduction of background emission. By using n-type-only gating and current-blocking barriers, the design mitigates p-dopant diffusion in InP and preserves field uniformity, providing a stable and scalable C-band platform critical for quantum photonic devices.

\begin{figure}[t]
    \centering
    \includegraphics[width=0.95\linewidth]{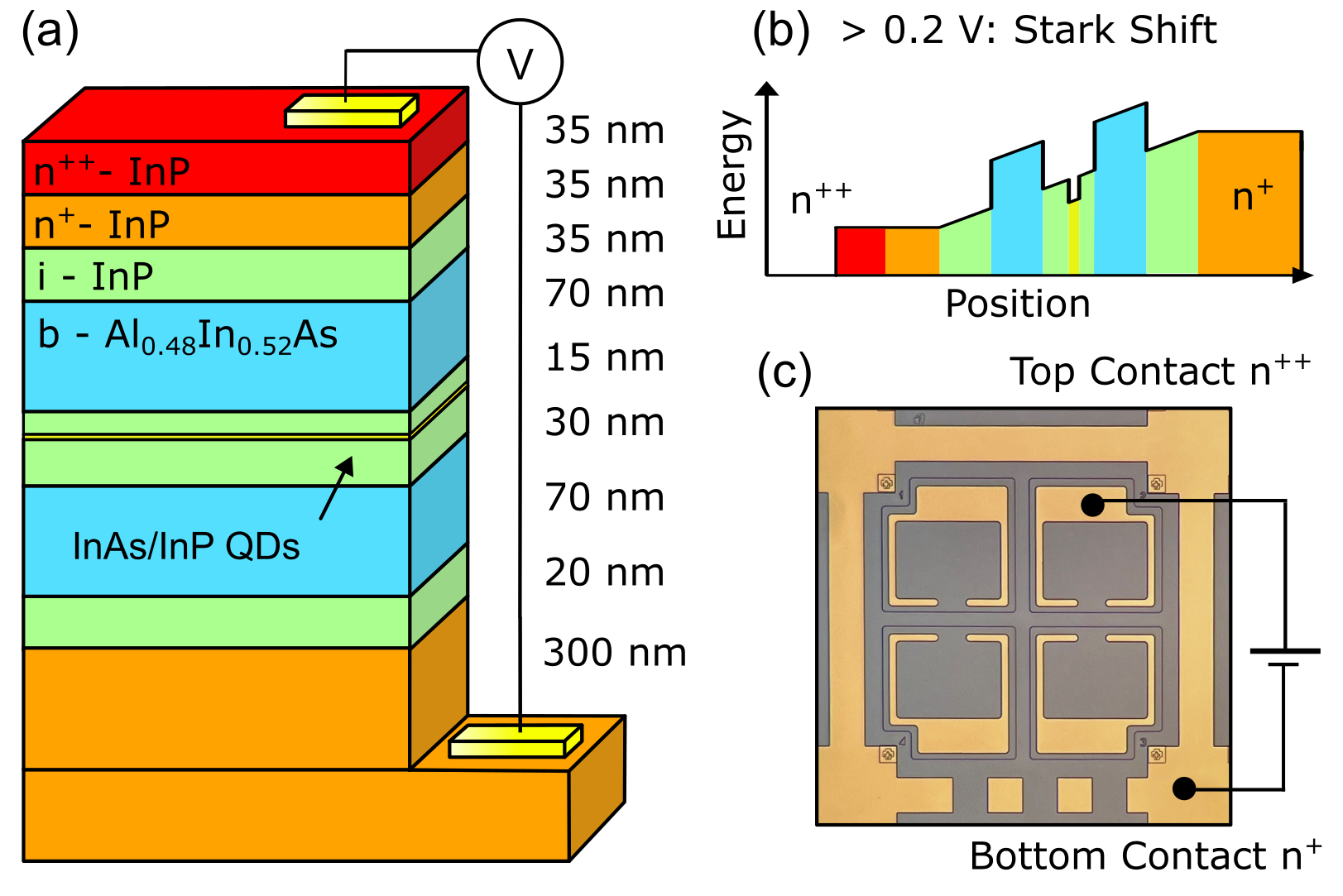}
   \caption{(a) Schematic of the sample structure. We begin with a 300\,nm n-doped InP layer 
    \(\bigl(2.0\times 10^{18}\,\mathrm{cm}^{-3}\bigr)\). 
    Next, a 70\,nm Al\(_{0.48}\)In\(_{0.52}\)As barrier is grown to reduce current and block carriers. The intrinsic region then comprises a 35\,nm undoped InP spacer, a 1\,nm droplet-epitaxy InAs quantum dot layer (for telecom-wavelength emission), and a 15\,nm undoped InP capping layer. Another 70\,nm Al\(_{0.48}\)In\(_{0.52}\)As barrier provides electrical isolation. Finally, a 35\,nm n\(^+\) InP layer \(\bigl(2.0\times 10^{18}\,\mathrm{cm}^{-3}\bigr)\) and a 35\,nm 
    n\(^ {++}\) InP layer \(\bigl(1.0\times 10^{19}\,\mathrm{cm}^{-3}\bigr)\) is deposited. (b) Schematic conduction band profile above 0.2 V Gate voltage. (c) Microscope image of fabricated mesa structure (0.14~$\mathrm{mm}^{2}$), showing the top and bottom contacts. }
    \label{fig:intro}
\end{figure}
\begin{figure*}[t]
    \centering
    \includegraphics[width=0.95\linewidth]{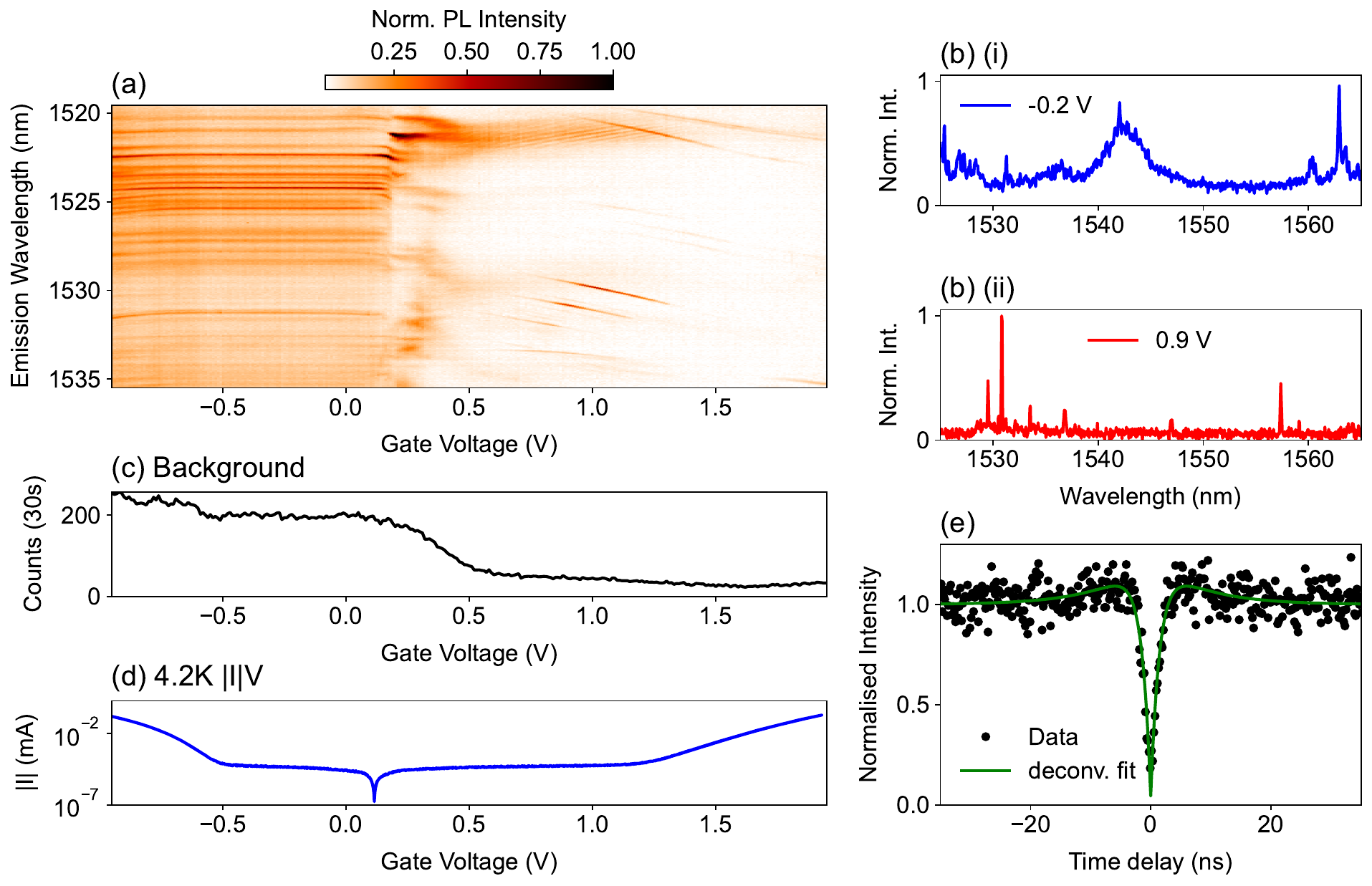}
    \caption{ (a) PL for 80 $\mu$W excitation  as a function of applied bias voltage for a QD, showing voltage dependent tuning of QD lines in C-band  (b) Representative normalized PL spectra from (a) for (i) high background -0.2 V bias voltage and (ii) low background 0.9 V gate voltage. (c) Average background count level with above band  excitation of the sample at 80 $\mu$W power at different gate voltages (d) $|I|  V$ characteristics of the measured device at 4.2 K (with 30 $\mu$W of above-band (852 nm) excitation power). (e) HBT measurement confirming high purity single-photon emission from the QD at a gate voltage of 1.18 V and emission wavelength of 1530.3~ nm. The normalized second-order correlation function \(g^{(2)}(\tau)\) exhibits a clear anti-bunching dip at \(\tau = 0\). Fitting yields \(g^{(2)}(0) = 0.04 \pm 0.04\). }
    \label{fig:background}
\end{figure*}
\begin{figure*}[t]
    \centering
    \includegraphics[width=0.98\linewidth]{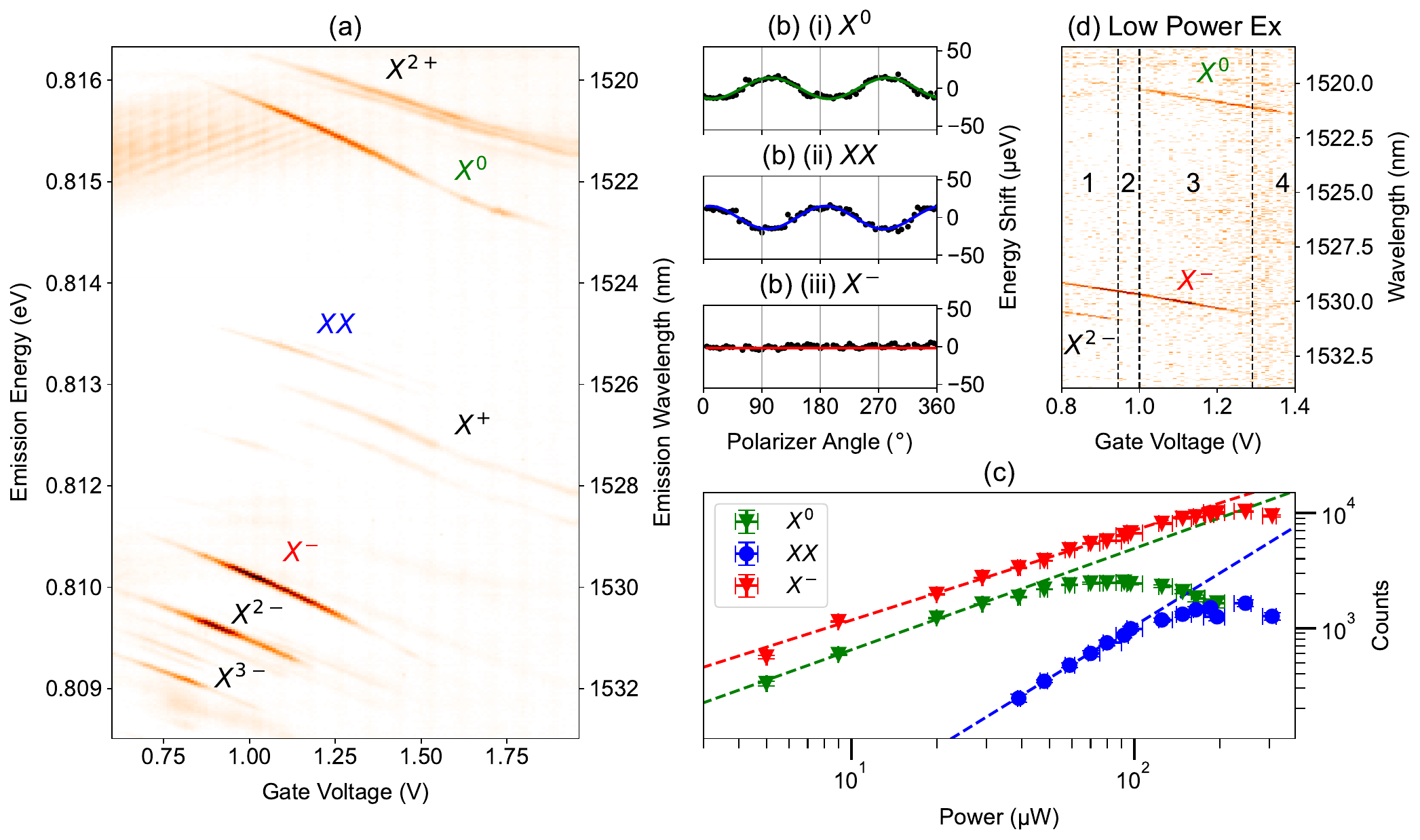}
    \caption{(a) Micro-PL of a single selected QD as a function of gate voltage, using an above-barrier pumping scheme, in the low background gate voltage regime. (b) Polarization dependent PL of the (i) $X^{0}$, (ii) XX and the $X^{-}$ (iii) emission lines at 1.1 V gate voltage. The emission lines in (i-ii) exhibit polarization-dependent shifts arising from their intrinsic fine-structure splitting, their converse polarization dependence confirm a direct radiative cascade. (c) PL intensity as a function of excitation power for the  $X^{0}$, $XX$ and $X^{-}$ states. The dashed lines represent the power-dependent fit before dot saturation.  (d) PL intensity spectra under weak above-band pumping (7\% of saturation) of the QD for different applied gate voltages. For specific voltage regions, 2 and 4, only the $X^{0}$ or the $X^{-}$ are present respectively.}
    \label{fig:charge_control}
\end{figure*}

The layer structure of the device studied here, shown in Fig \ref{fig:intro}(a), began with the deposition of a 300~nm n-doped InP layer (\(2.0 \times 10^{18}~\mathrm{cm^{-3}}\)) at 610~$^\circ$C. This was followed by 20~nm of undoped InP. A 70~nm Al\(_{0.48}\)In\(_{0.52}\)As layer lattice-matched to InP was then deposited, serving as a barrier and current blocking layer, allowing current management and carrier confinement within the active region. Then a 30~nm undoped InP spacer, a central layer of InAs QDs, and a 15~nm undoped InP capping layer was grown.  The QDs were grown by DE in MOVPE via the deposition of In droplets, subsequently crystallized by an As flow. For further details on the growth sequence, we refer to our previous work \cite{sala2020, sala2021}.  The intrinsic region was then followed by another 70~nm Al\(_{0.48}\)In\(_{0.52}\)As layer that again acts as a barrier and current blocking layer, and a further 35 nm of undoped InP.  The structure was finalized with a 35~nm n\(^ {+}\) InP layer (\(2.0 \times 10^{18}~\mathrm{cm^{-3}}\)) followed by a 35~nm n\(^ {++}\) InP contact layer (\(1.0 \times 10^{19}~\mathrm{cm^{-3}}\)). Fig.\ref{fig:intro}(b) presents a schematic diagram of the conduction band at a gate bias of +0.2 V for the structure (see Supplementary section S2 for calculated 1D band structures). Fig.\ref{fig:intro}(c) shows an optical image of the fabricated 0.14~$\mathrm{mm}^{2}$ mesa struture, which highlights the position of the top and bottom contacts.

The QD PL intensity presented in Fig.\ref{fig:background}(a) shows the PL from the wafer under above‐band (852 nm) excitation as a function of gate voltage applied. In Fig.\ref{fig:background}(b), representative spectra from the data presented in (a) at –0.2 V (panel i) exhibit pronounced background emission, whereas at 0.9 V (panel ii) the background is significantly suppressed. We note that the pronounced asymmetry and persistent background emission observed in Figs.\ref{fig:background}(a–b) may arise from charge accumulating in or around the intrinsic region, which alters the local band profile and leads to an asymmetric, broadened background continuum. Fig.\ref{fig:background}(c) shows that the average background count rate (over the range 1518.5-1565 nm) under above‐band (852 nm) excitation decreases substantially as the gate voltage increases. The reduction in background emission at higher gate voltages correlates with changes in current through the device, as shown by the I–V characteristics shown in Fig.\ref{fig:background}(d), measured at 4.2 K under 30 $\mu$W excitation (see Supplementary Section S3 for a more detailed discussion of the I–V characteristics). Using a Hanbury Brown–Twiss setup under CW excitation at 852 nm with a gate voltage of 1.18 V and emission at 1530.3 nm, we confirm the single‐photon character of the QDs’ telecom‐wavelength emission. As shown in Fig.\ref{fig:background}(e), we observe a raw coincidence minimum of $g^{(2)}(0)\approx0.18$. Deconvolving the $g^{(2)}(\tau)$ curve (accounting for the finite bin width and instrument response time) yields $g^{(2)}(0) = 0.04 \pm 0.04$. For the same state at the same voltage, time-resolved photoluminescence measurements under pulsed excitation at 830 nm yielded a lifetime of 2.2 ns (see Supplementary Section S4).

Gate voltages not only influence background emission but also enable controlled loading of individual QDs with electrons. Fig. \ref{fig:charge_control}(a) shows the QD PL, mapped as a function of $V_{\mathrm{gate}}$ and the emission energy, revealing transitions between the neutral excitonic species ($X^0$) and singly charged ($X^{-}$) excitonic species. To verify the charge states and excitonic behavior, we measured both the FSS and power dependence for each highlighted state in Fig \ref{fig:charge_control}(a). Neutral excitonic species exhibit FSS due to exchange interactions between electron and hole spin states \cite{PhysRevB.65.195315}. In contrast, the additional electron in $X^-$ suppresses this interaction and therefore the emission line displays no FSS. The presence and degree of FSS was identified through polarization-dependent measurements, in which a linear polarizer in the collection path was rotated. Fig \ref{fig:charge_control}(b) shows the results of these measurements, where for a range of angles of the linear polarizer ($0-360^{\circ }$), the energy of the intensity peak of the PL was determined. The FSS was then calculated as the energy separation between the minimum and maximum energy across the range of angles. Fig \ref{fig:charge_control}(b, i-ii) show FSS indicating neutral states, while (iii) shows no appreciable FSS indicating the emission line corresponds to a charged state. Considering the relative binding energies of the states \cite{peniakov2025}, (iii) was determined to be a negatively charged trion ($X^{-}$).  Additionally, we used power dependent PL measurements to identify single and biexcitonic emission lines. Because biexciton formation requires two e–h pairs, its PL intensity increases roughly twice as much as a single exciton for the same increase in excitation power \cite{kettler2016}. Fig \ref{fig:charge_control}(c) shows the PL intensities of each state at excitation powers ranging from 5\,\textmu W to 360\,\textmu W (using an 850\,nm continuous wave (CW) diode laser). Each peak was fit to determine the maximum PL as a function of external applied excitation power. The extracted slopes of: 0.78, 0.88, and 1.51 for $X^{-}$, $X^{0}$, and $XX$, respectively, indicating a $\thicksim1.8$ greater slope for $XX$ compared to $X^{0}$.

The application of an external electric field enables the continuous, reversible tuning of the emission wavelength of QDs~\cite{PhysRevLett.93.217401,Nowak2014}. The tuning range for the three exciton states (operating within a signal-to-background ratio of 5:1), was determined by sweeping the gate voltage from 0.59\,V to 1.96\,V. We extract the tuning ranges of 2.40 nm, 0.82 nm and 1.73 nm for the $X^{0}$, $X^{-}$ and $XX$ lines, respectively (see Section supplementary section S1). Such spectral control is indispensable for integrating QD emitters into scalable photonic networks, where frequency matching across multiple sources is crucial~\cite{photonic_qi_review}.

The ability to control the QD charge state is important in understanding the QD's level structure and has practical applications in spin-based quantum information processes \cite{ediger2005}. Charge control enables precise initialization and readout of spin states \cite{heindel2023}, the foundational elements for spin-photon interfaces \cite{Warburton2013}. To clearly distinguish each excitonic transition in the PL, we employed a low-power, above-band optical excitation. Under these conditions, each charge configuration emerges in different gate voltage ranges, avoiding higher power phenomena such as substantial space-charge build-up or excessive spectral broadening \cite{pinin, Warburton2013, arakawa2020}. The PL as a function of the gate voltage in this weak excitation regime is shown in Fig.~\ref{fig:charge_control}(d), where four gate-voltage regimes can be distinguished: in Region 4 (1.3–1.4 V) the QD is neutral and emits only the $X^{0}$ line; in Region 3 (1.0–1.29 V) both $X^{0}$ and $X^{-}$ transitions appear; in Region 2 (0.945–1.0 V) the $X^{-}$ line dominates as the QD holds an excess electron, suppressing $X^{0}$; and in Region 1 (0.8–0.945 V) the doubly charged exciton $X^{2-}$ is observed, indicating higher electron occupancy. These regions correspond to the sequential charging of the QD through the Coulomb blockade regime, enabling clear identification of each excitonic transition under weak excitation~\cite{pinin,Warburton2013,arakawa2020}. 

\begin{figure}[t]
    \centering
    \includegraphics[width=0.98\linewidth]{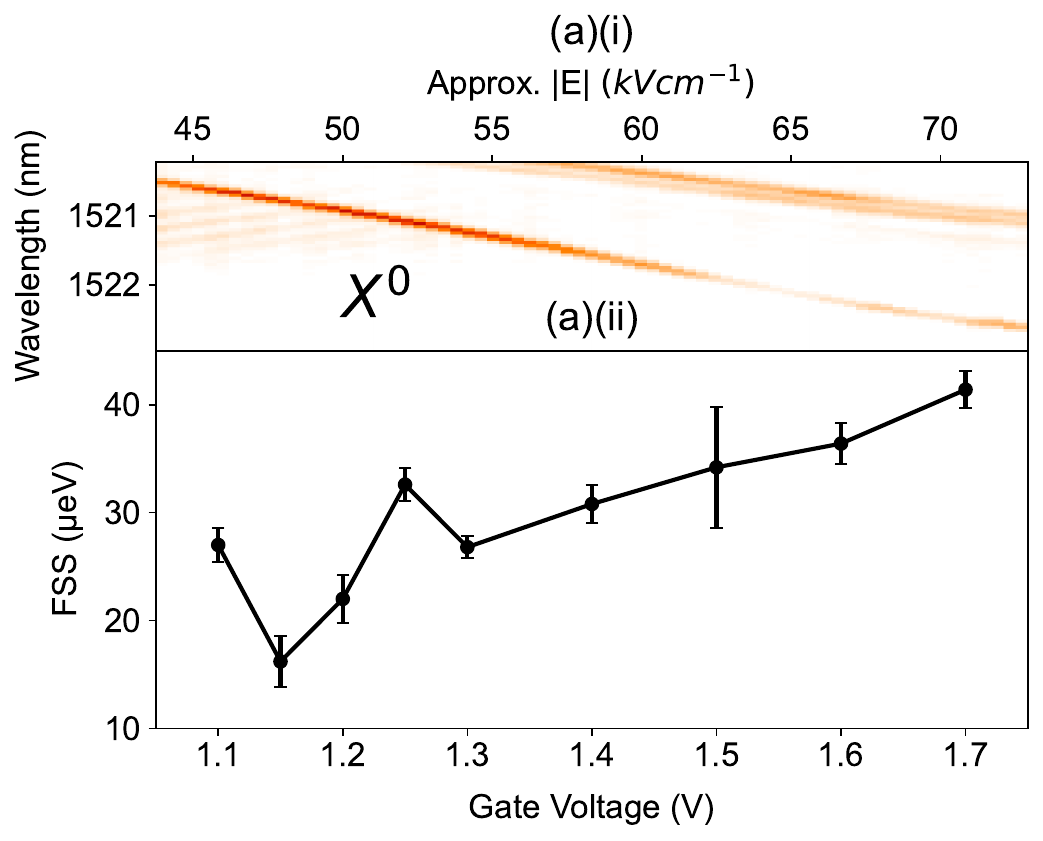}
    \caption{(a)(i) PL Intensity of the $X^{0}$ state and (ii) Measurements of the fine structure splitting as a function of the applied gate voltage.}
    \label{fig:fss_voltage}
\end{figure}

Moreover, in the QD tuning range, the application of the electric field is also capable of reducing the FSS \cite{bennett2010}, a critical capability to preserve high-fidelity polarization entanglement in the biexciton–exciton cascade \cite{he2008,laccotripes2023, anderson2020}. The FSS results from breaking the QD symmetry, in physical shape, composition, or strain terms \cite{he2008}. Since such asymmetries are ubiquitous, in situ FSS-tuning methods are essential for realizing high-fidelity entangled-photon sources. In our device, an applied electric field modifies the exciton wavefunctions via the quantum-confined Stark effect, reducing the effective anisotropic exchange and thus the FSS \cite{Ghali2012}. Other established methods for controlling the FSS include the application of stress \cite{seidl2006,trotta2016,Millington-Hotze2024} and magnetic fields \cite{stevenson2006}. To demonstrate electrical control over the degree of FSS in our system, we performed polarization-resolved PL spectroscopy to measure the QD's FSS as a function of voltage. Fig.~\ref{fig:fss_voltage}(i) shows the $X^{0}$ PL intensity as a function of gate voltage, with (ii) showing the corresponding measurement of the FSS as a function of gate voltage. Additionally the magnitude of the electric field is estimated for a $d_{i}$= 240~nm intrinsic region ($E=V/d_{i}$), across this voltage range. The extracted FSS ranges from $41 \pm 2 \,\mu eV$ at 1.7~V to $16 \pm 2 \, \mu eV$ at 1.15~V. This demonstrates electrical control of the FSS in droplet epitaxy QDs at the telecom C-band. While an FSS of $16 \pm 2 \, \mu eV$ is not itself sufficient to ensure high-fidelity entanglement, the ability to tune the FSS electrically provides a critical control ‘dial’ that, in combination with the pre selection of QDs with naturally small FSS and strain-tuning techniques \cite{lettner2021}, underscores the viability of this system for the development of telecom wavelength entangled SPS.


\label{sec:discussion}
In this work, we have introduced and validated an $n^{++}\!-\!i\!-\!n^{+}$ InP diode containing droplet epitaxy InAs QDs, demonstrating tunable single-photon generation at telecom wavelengths. Our results establish this structure as a promising platform for key quantum technologies. A principal advantage of this n-type gating scheme is its ability to mitigate p-dopant diffusion, a common issue in InP-based devices, thus opening the way to stabilizing the charge environment around each QD. By generating photons in the telecom C-band, future devices will be able to leverage the existing fiber infrastructure for practical quantum communication \cite{northup2014}, while reduced fiber attenuation at C-band specifically benefits secure data links and distributed quantum computing. 

The observed Stark shifts exceeding 2.4~nm confirm a useful single QD emission tuning range. Reliable tuning of a QD’s emission wavelength is crucial for several key functionalities in quantum photonics. First, by finely adjusting the QD’s transition into resonance with a high quality factor cavity mode, one can exploit the Purcell effect to achieve ultrafast, deterministic single‐photon emission  \cite{purcell_brash}, directional emission \cite{chiral_cav} and, in the strong‐coupling regime, engineer zero‐dimensional polaritons with well‐defined energy splitting and lifetimes \cite{Laucht_2009,PhysRevLett.104.047402}. Second, precise spectral alignment between two spatially separated QDs enables coherent dipole‐dipole coupling \cite{controlling_coherence,lodahl_super} or the emission of indistinguishable photons for entanglement protocols, both of which demand sub‐meV tuning accuracy to overcome inhomogeneous broadening and fabrication variability. Finally, matching the QD’s emission to narrow atomic transitions (e.g., the ORCA memory at 1529.3 nm \cite{ORCA}) allows deterministic storage and retrieval of single‐photons in atomic ensembles \cite{ORCA}, thereby forming a hybrid quantum node that combines solid‐state emitters with long‐lived atomic memories.

Furthermore, robust control over $X^0$ and $X^{-}$ states highlights the potential of the device for spin-photon interfaces \cite{lodahl2015,kuhlmann2013,somaschi2016}. The demonstrated charge control opens possibilities for spin-based memory \cite{sun2018,dreiser2008} and quantum logic \cite{DeSantis2017}, where spin pumping \cite{carter2013} and coherent spin readout \cite{antoniadis2023} at C-band could support quantum gates and small-scale entanglement on the same platform. Our demonstration of tunable single-photon emission in the telecom band thus marks a key milestone toward scalable quantum photonic circuits, and continued integration efforts and system-level demonstrations will further unlock the potential of these devices for quantum communication and computation at telecom wavelengths.

See the Supplementary Material for additional experiments, analyses, and simulations supporting the main study. S1 details extraction of the QD tuning range from photoluminescence (PL) spectra; S2 reports band-structure simulations of the studied heterostructure; S3 provides low-temperature I–V characteristics across excitation powers and temperatures; S4 presents time-resolved PL and the lifetime of the $X^{-}$ line.

This work was supported by EPSRC Grant No. EP/V026496/1 and the Integrated Quantum Network Hub EP/Z533208/1. A.J.B. acknowledges additional support from the EPSRC (UK) fellowship EP/W027909 and the Royal Society Research Grant RG/R2/232470. The authors acknowledge helpful discussions with Dr Andrew Foster and Dr Mahmoud Jalali Mehrabad.

\section{Author Declarations}
\subsection{Conflict of Interest}
The authors have no conflicts to disclose.
\subsection{Author Contributions}
N.J.M. conceived the project, designed the electronic structure of the wafer, and together with A.J.B. oversaw the experimental work. N.J.M., A.J.B., A.T., C.L.P., L.H., P.M-H., E.O.M., and D.H. carried out the experimental measurements. E.M.S. grew the QD wafer. R.D. performed device fabrication. E.O.M and K.A.O carried out the band structure simulations. J.H., M.S.S., A.M.F and L.R.W. provided supervision and expertise. N.J.M., A.T. and E.O.M. wrote the manuscript with input from all authors.

\section{Data Availability}
The data that support the  findings of this study are available from the corresponding author upon reasonable request.

\bibliography{Bibli}

\end{document}